\documentclass[12pt]{article}
\title{
Nonperturbative corrections to the quark self-energy}
\author{Yu.A.Simonov \\
State Research Center,\\ Institute of Theoretical and Experimental
Physics, Moscow, Russia}
\date{}
  \newcommand{\be}{\begin{equation}}
\newcommand{\ee}{\end{equation}}  
\def\fun#1#2{\lower3.6pt\vbox{\baselineskip0pt\lineskip.9pt
\ialign{$\mathsurround=0pt#1\hfil ##\hfil$\crcr#2\crcr\sim\crcr}}}

\newcommand{\vesig}{\mbox{\boldmath${\rm \sigma}$}}
\newcommand{\vep}{\mbox{\boldmath${\rm p}$}}

\newcommand{\veB}{\mbox{\boldmath${\rm B}$}}
\newcommand{\veE}{\mbox{\boldmath${\rm E}$}}

 \newcommand{\lan}{\langle}
 \newcommand{\ran}{\rangle}
\begin{document}

\maketitle

\begin{abstract}
Nonperturbative  self-energy  of a bound quark is computed
gauge-invariantly in the framework of background perturbation
theory. The resulting $\Delta m^2_q$ is negative and is a
universal function of string tension and current quark mass. The
shift of hadron mass squared, $M^2$, due to $\Delta m^2_q$ is
negative, large for light quarks, and solves the long-standing
problem of Regge intercepts in relativistic quark models.

\end{abstract}

\section{Introduction}

The quark self-energy (QSE) operator  for a quark propagating
through the vacuum background field can be introduced in a
gauge-invariant way, when quark is a part of a white object. Then
one associates QSE with a part of the total energy of the bound
state which does not depend on characteristics of other
constituents, but depends on the given quark motion. In contrast
to the perturbative QSE which is gauge and renormalization-scheme
dependent \cite{1} one can introduce the nonperturbative QSE in
the framework of the Background Perturbation Theory  (BPTh)
\cite{2,3}, in which case QSE can be expressed through the  gauge
invariant field  correlators  (for a  general formalism of field
correlators see \cite{4,5}).

  It is useful to separate the nonperturbative  contributions to QSE into two
  parts; one, coming from the color quark charge interaction with
  the nonperturbative  background,  and another created by the color magnetic
  moment of the quark (analogs of diamagnetic and paramagnetic
  contributions).

  It was shown in \cite{4,5} that the quark charge part is responsible for
  creation of confining  interaction hence is nonlocal and should
  be attributed to the $q\bar q$ interacting kernel. Only small
  distance-independent corrections can be separated out of this
  part and associated  either with QSE or with $q\bar q$ interaction.

  The color-magnetic moment part of QSE is instead large and
  independent of the string and antiquark motion (in the limit of
  small correlation length $T_g$ of field correlators).
 It gives the dominant part of QSE and has  specific negative
 sign, decreasing the hadron mass. In what follows we shall
 compute only this part and shortly discuss corrections from the
 charge  part in the conclusions.

  The next step is to calculate the contribution of QSE to the
  hadron mass and Regge trajectories. To this end one can use the
  Feynman-Schwinger (world-line) representation \cite{6}-\cite{9'} and
  following from it the Hamiltonian einbein technic, introduced in
  \cite{9,10} (for reviews see lectures \cite{9',11} and recent paper
  \cite{12}).
    As  a
  result one  obtains a negative self-energy contribution from
  each quark and antiquark depending only on string tension and
  current quark mass.

  Taken as a correction to the total hadron mass, this selfenergy
  contribution is  favoured from phenomenological point of view.
  First of all it explains
   sign and magnitude
  of  negative constant which is always  added to the relativistic quark
  Hamiltonian in most existing models.  In a similar way is
  resolved the old problem of Regge intercepts $J(M^2=0)$, when intercepts
  come out too low when computed through string tension without
  introduction of that  phenomenological constant.
  Some examples for meson and baryon masses are given and the net
  effect for the Regge slope is discussed. The nontrivial
  character of the hadron mass shift obtained below is that
  it corrects  the Regge intercept and keeps at the same time
   the value of the Regge
  slope at the standard string value $2\pi\sigma$. (One should
  note that simply adding a constant to the hadron mass, as it is
  usually done phenomenologically, would spoil the universality of
  Regge slope).

  The plan of the paper is as follows. In section 2 the quark
  Green's function in the framework of  BPTh is introduced and the
  world-line integral is used to separate contributions of color
  charge and color magnetic moment of the quark. In section 3 the
  lowest order (Gaussian) contribution to QSE is written down and
  expressed through the Gaussian correlators. In section 4
  self-energy corrections to the Hamiltonian are derived and
  estimated, and the problem of Regge intercepts is discussed,
  while section 5 is devoted to conclusions and outlook.

  \section{Quark Green's function in vacuum background fields}

Gluons play two different roles in QCD, namely, they create gluon
condensate and string  between quark and antiquark on one hand,
and also  propagate in the nonperturbative vacuum as valence
gluons. Therefore it is convenient following \cite{2,3} to
separate the total gluonic field $A_\mu$ into background NP field
$B_\mu$ and field of valence gluons $a_\mu$,
\be
A_\mu= B_\mu+ a_\mu \label{1} \ee and using the 'tHooft identity
to write the QCD partition function as $$ Z(J) =\frac{1}{N} \int
e^{-S_E(A)+\int J_\mu A_\mu dx} DA D\psi D\bar \psi=$$
\be
=\frac{1}{N'}\int DB \eta (B) e^{JBdx} \int Da D\psi D\bar \psi
e^{-S_E(B+a)+\int J_\mu a_\mu dx} \label{2}\ee with $N'=N\int
DB\eta (B)$. Here $\eta(B)$ is  an arbitrary measure  and can be
put $\eta(B)\equiv 1$ or chosen in such a way as to improve
convergence of perturbation series in $ga_\mu$. For what follows
the value of $\eta(B)$ and explicit details of  averaging over
fields $B_\mu$ are not important.

According to (\ref{2}) one can write any vacuum average as
\be
\lan F(A)\ran_A=\lan\lan F(B+a)\ran\ran_{B,a} \label{3} \ee

The Euclidean quark Green's function $G_q(x,y)$ can be  written
using the Feynman-Schwinger (world-line) Representation (FSR)
\cite{6}-\cite{8} as $$ G_q(x,y)= (m_q+\hat D)^{-1}_{x,y}=
(m_q-\hat D)_x(m^2_q-\hat D^2)^{-1}_{x,y}= $$
\be
=(m_q-\hat D)_{x}\int^\infty_0 ds (Dz)_{xy} e^{-K} P_A\exp (ig
\int^x_y A_\mu dz_\mu) P_F\exp ( \int^s_0 g\sigma F d\tau),
\label{4}\ee where $P_A, P_F$ are ordering operators of matrices
$A_\mu$ and $F_{\mu\nu}$ respectively, $\sigma F\equiv
\sigma_{\mu\nu} F_{\mu\nu}=\left( \begin{array}{ll} \vesig\veB&
\vesig \veE\\ \vesig \veE& \vesig \veB\end{array} \right),$ and
$\sigma_{\mu\nu}=\frac{1}{4i} (\gamma_\mu\gamma_\nu-\gamma_\nu
\gamma_\mu),~~ K=m^2_qs+\frac14 \int^s_0 \dot z^2_\mu d\tau$.

In (\ref{4})  the quark moving along its trajectory $z_\mu(\tau)$
is interacting with external total field $A_\mu$ by its color
charge (the first exponent) and by its color  magnetic moment (the
second exponent in (\ref{4})). As we shall see it is the second
interaction which yields the most important effect in creating the
nonperturbative mass shift  of the quark, and we shall calculate
it in terms of field correlators. One should also note that the
term $ga_\mu$ in $\hat D=\gamma_\mu(\partial_\mu-ig a_\mu-ig
B_\mu)$ produces usual perturbative quark mass correction
(including anomalous dimension for the mass evolution) and was
investigated  thoroughly especially for heavy quarks , (see
\cite{13} and references therein),  and we shall not enter in
discussion of this topic. Instead we shall concentrate on the
contribution of nonperturbative background $B_\mu$.

The Green's function (\ref{4}) is not gauge-invariant. To define
quark mass corrections in a gauge-invariant way, one must  use the
Green's function  of  a white object --  of a meson or baryon. In
the former case one can write \cite{14} (in the flavour-nonsinglet
case and neglecting sea-quark determinant) $$ G_M(x,y)=\lan tr
\Gamma_1 G_q(x,y) \Gamma_2 G_{\bar q} (x,y)\ran_B=$$ $$=\lan tr \{
\Gamma_1 (m_{q_1}-\hat D) \int^\infty_0 ds_1 Dz\int^\infty_0 ds_2
D\bar z e^{-K_1-K_2} \Phi_\sigma(x,y)\times $$ \be\times
\Gamma_2(m_{q_2}-\hat D) \Phi_\sigma(y,x)\}\ran_B.
 \label{5} \ee

 Here $\Phi_\sigma(x,y)$ is the product of the last two exponents
 on the r.h.s. of (\ref{4}), where  the field $A_\mu$ is replaced by
 $B_\mu$ (setting $a_\mu\equiv 0)$: $ \Phi_\sigma (x,y)= P_BP_F\exp
 [ig\int^x_y B_\mu dz_\mu+ \int^s_0 g\sigma F d\tau].$

 One can note that the integrand of (\ref{5}) contains the closed
 Wilson loop with insertions of operators $\hat D$ (twice) and
 operators $\sigma F$ (infinitely many times). Therefore the whole
 construction in (\ref{5}) is gauge  invariant. The sign of trace
 in (\ref{5}) implies summing over Lorentz and color indices, and
 $\Gamma_i (i=1,2)$ stands for current vertices,
 $\Gamma_i=1,\gamma_5,\gamma_\mu,...$

 \section{Quark self-energy correction due to nonperturbative
 background}

 Consider the FSR for the  quark Green's function (\ref{4}). In the 2-nd order
  of  perturbative expansion
it can be written as $$ G_q(x,y) = (m_q-\hat D) \int^\infty_0 ds
\int^s_0 d\tau_1 \int^{\tau_1}_0 d\tau_2 e^{-K} (Dz)_{xu} d^4u
(Dz)_{uv} d^4 v(Dz)_{ vy}\times $$ \be (ig A_\mu(u)\dot
u_\mu+g\sigma_{\mu\nu} F_{\mu\nu} (u))  (ig A_\nu(v)\dot v_\nu+
g\sigma_{\lambda\sigma} F_{\lambda\sigma} (v)), \label{6}
 \ee
 where we have used the identities
\be
 (Dz)_{xy} =(Dz)_{xu} d^4u(Dz)_{uv} d^4 v(Dz)_{vy},
 \label{7}
 \ee
 \be
 \int^\infty_0ds\int^s_0 d\tau_1\int^{\tau_1}_0 d\tau_2 f(s,\tau_1,
 \tau_2) =
\int^\infty_0ds\int^\infty_0 d\tau_1\int^\infty_0 d\tau_2
f(s+\tau_1+
 \tau_2, \tau_1+\tau_2,\tau_2)
 \label{8}
 \ee
At this point one can expand the quark Green's function only in
color magnetic moment interaction ($\sigma F$), which is useful
when spin-dependent interaction can be treated perturbatively, as
it is in most cases for mesons and baryons \cite{11} ( the
exclusions are Goldstone bosons and nucleons, where spin
interaction is very important and interconnected with chiral
dynamics). In this case to the second order in ($\sigma F$) one
obtains $$ G_q^{(2)}(x,y) =(m_q-\hat
D)\int^\infty_0ds\int^\infty_0 d\tau_1\int^\infty_0 d\tau_2
e^{-m^2_q(s+\tau_1+\tau_2)-K_0-K_1-K_2} (Dz)_{xu}\times $$ \be
\times \Phi (x,u) g(\sigma F(u))  d^4 u(Dz)_{uv} \Phi(u,v)
g(\sigma F(v))  d^4v(Dz)_{vy}. \label{9} \ee

In another way  expanding $G_q= (m_q-\hat D)(m^2_q-D_\mu^2-g\sigma
F)^{-1}$ in powers of $ (g\sigma F)$ it can be written as $$
G^{(2)}_q(x,y) =(m_q-\hat D) (m^2_q-D^2_\mu)^{-1}_{xu}
d^4u~g(\sigma F(u))(m^2_q-D^2_\mu)^{-1}_{uv} d^4v\times $$ \be
\times g(\sigma F(v)) (m^2_q-D^2_\mu)^{-1}_{vy}. \label{10}
 \ee

One can easily find the correspondence between (\ref{9}) and
(\ref{10}) since $(m^2_q-D^2_\mu)^{-1}$ is the scalar Green's
function having the FSR as follows
\be
G(x,y)\equiv (m^2_q-D^2_\mu)^{-1}_{xy}=\int^\infty_0 ds(Dz)_{xy}
e^{-K} \Phi_\sigma (x,y).\label{11}\ee

Now keeping terms $g(\sigma F)$ up to the second order one can
find from (\ref{10}) and (\ref{5}) (leaving aside for simplicity
the factors $(m_q-\hat D)$ in (\ref{5})) that one has in (\ref{5})
an expansion
\be
tr\lan W\ran_B+tr\lan P_FW g\sigma F(u) g\sigma F(v)\ran_B+...
\label{12}\ee where operators $g\sigma F$ are inserted inside
Wilson loop in the properly  ordered form  due to the operator
$P_F $, keeping gauge invariance.

Using cluster expansion for $W$ \cite{4}   the last term in
(\ref{12}) was computed in the Appendix of \cite{14} with the
result $$ \lan tr P_F F_{\mu\nu}(u) F_{\rho\lambda}(v) W(C)\ran =
tr\{[\lan F_{\mu\nu} (u, x_0) F_{\rho\lambda} (v, x_0)\ran- $$ \be
-g^2\int d\sigma_{\alpha\beta} (w) \lan F_{\mu\nu} (u, x_0)
F_{\alpha\beta} (w, x_0)\ran \int d\sigma_{\gamma\delta}(z)\lan
F_{\rho\lambda} (v, x_0)_{\gamma\delta}  F(z,x_0)\ran ] \lan
W(C)\ran\} \label{13}\ee
 Here $F(x,x_0)=\Phi(x_0, x) F(x) \Phi(x,x_0)$, with $\Phi(x,y)
 \equiv  P\exp ig \int^x_y B_\mu (z) d z$. It is convenient in
 (\ref{13}) to choose the point  $x_0$ (note that $\lan W(C)\ran$ does not
 depend on $x_0$) on the line between points $u$ and $v$, since in
 this case the first correlator in (\ref{13}) takes the form,
 studied in \cite{4}

$$ g^2\lan F_{\mu\nu} (u) \Phi( u, v) F_{\rho\lambda} (v)
\Phi(v,u)\ran= $$ $$ =\hat 1_{ab}\{ (\delta_{\mu\rho}
\delta_{\nu\lambda} -\delta_{\mu\lambda}- \delta_{\mu\lambda}
\delta_{\nu\rho}) D(u-v) +\frac12 [\partial_\mu
(h_\rho\delta_{\mu\lambda}-h_\lambda\delta_{\nu\rho})+$$
\be
+\partial_\nu(h_\lambda\delta_{\mu\rho}-h_\rho\delta_{\mu\lambda})]D_1(u-v)\},~~
h_\mu\equiv u_\mu-v_\mu,\label{14} \ee where $\hat 1_{ab}$ is the
unit operator in color space.

In what follows we shall keep only the first term in  Eq.
(\ref{13}) in  the   form ((\ref{14}), since the second term
describes interaction of the quark spin with the  string world
sheet between the quark and antiquark and cannot be clearly
associated with self-energy (actually this term contributes to the
spin-orbit interaction between quark and antiquark, see discussion
in \cite{14,15}).

Multiplying (\ref{13}), (\ref{14}) with  spin operators one has
\be g^2\lan \sigma F(u) \Phi(u,v) \sigma F(v) \Phi(v,u)\ran =\hat
1_{ab}\hat 1_{\alpha\beta} \{8 D(u-v)+\frac32\partial_\nu (h_\nu
D_1(u-v))\}.\label{15}\ee

Here $\hat1_{\alpha\beta}$ is the unit operator in bispinor
indices.

>From (\ref{10}) one can see that the following combination occurs
playing the role of quark  selfenergy correction
\be
\Lambda\equiv \int d^4(u-v) g^2\lan \sigma F(u)\Phi(u,v)\sigma
F(v)\Phi(v,u)\ran G(u,v)\label{16}\ee where
$G(u,v)=(m^2_q-D^2_\mu)^{-1}_{u,v,}$ is given in (11). Since
$G(u,v)$ depends on the interaction of quark with the string world
sheet (through $D_\mu=\partial_\mu-ig B_\mu$), then in general the
correction (\ref{16}) cannot be unambiguously attributed to QSE.
However, at this point one can use the important fact, that $D(u)$
and $D_1(u)$ are short-ranged functions with the gluonic
correlation length $T_g\approx 0.2 \div 0.3 $ fm from lattice
calculations in the cooled vacuum \cite{16} and even smaller, if
one extracts $T_g $ from  the gluelump masses \cite{17}. Therefore
integration in (\ref{16}) is limited to small $|u-v|$ and in this
region one can replace $G(u,v)$ by $G_0(u,v)$ (since $G_0$ yields
singular terms $\sim \frac{1}{(u-v)^2}$, while background field
corrections give nonsingular contributions of the order of
$T^2_g\lan F^2\ran$. Thus one can expect the accuracy of this
replacement to be of the order of $g^2T_g^4\lan E^2_i\ran \sim$
10\%). Replacing $G$ by $G_0$ in (\ref{16}) and using for $D,D_1$
exponential form found on the lattice \cite{16}, one obtains in
this approximation from (\ref{16}) $$ \Lambda_0=\int d^4w\{ 8 D(0)
e^{-\delta|w|} +\frac32 D_1(0)
\partial_\nu (w_\nu e^{-\delta|w|})\}\times $$\be \times \frac{m_q
K_1(m_q|w|)}{4\pi^2|w|},~~ \delta\equiv \frac{1}{T_g}.
\label{17}\ee

It is convenient to define an integral $(K_1(x)$ is the Mc Donald
function)
\be
\varphi(m_q,\delta) =\int^\infty_0 w^2 dwe^{-\delta
w}K_1(m_qw)\label{18}\ee which has analytic form for $\delta>m_q$
\be
\varphi(m_q,\delta)=-\frac{3m_q\delta}{(\delta^2-m^2_q)^{5/2}}\ln
\frac{\delta+\sqrt{\delta^2-m^2_q}}{m_q} +
\frac{\delta^2+2m^2_q}{m_q(\delta^2-m^2_q)^2}.\label{19}\ee

In the opposite case, $\delta<m_q$, one has
\be
\varphi(m_q,\delta)=-\frac{3m_q\delta}{(m^2_q-\delta^2)^{5/2}}\arctan
\frac{\sqrt{m^2_q-\delta^2}}{\delta} +
\frac{\delta^2+2m^2_q}{m_q(\delta^2-m^2_q)^2}.\label{20}\ee

Two limiting cases are of interest for  what follows,
\be
\varphi(m_q,\delta)|_{m_q\to 0}=\frac{1}{m_q\delta^2}\label{21}\ee
\be
\varphi(m_q,\delta)|_{m_q\to\infty}=\frac{2}{m_q^3}-\frac{3\pi\delta}{2m^4_q}
+O\left(\frac{\delta^2}{m^5_q}\right).\label{22}\ee

In a similar way one can define the integral in front of $D_1(0)$
in Eq. (\ref{17}),  $$ \varphi_1(m_q,\delta)\equiv
(4+\delta\frac{\partial}{\partial\delta})\varphi(m_q,\delta)=$$
 $$=\frac{15
m^3_q\delta}{(\delta^2-m^2_q)^{7/2}}\ln
\frac{\delta+\sqrt{\delta^2-m^2_q}}{m_q} -
\frac{8m^4_q+9m^2_q\delta^2-2\delta^4}{m_q(\delta^2-m^2_q)^3};(\delta>m_q)$$
\be =-\frac{15 m^3_q\delta}{(m^2_q-\delta^2)^{7/2}}\arctan
\frac{\sqrt{m^2_q-\delta^2}}{\delta} +
\frac{8m^4_q+9m^2_q\delta^2-2\delta^4}{m_q(m^2_q-\delta^2)^3};
(\delta<m_q).\label{23}\ee The limiting values of $\varphi_1$ are
\be
\varphi_1(m_q,\delta)|_{m_q\to 0}
=\frac{2}{m_q\delta^2}\label{24}\ee

\be
\varphi_1(m_q,\delta)|_{m_q\to \infty}
=\frac{8}{m_q^3}-\frac{15\pi\delta}{2m^4_q}
+O\left(\frac{\delta^2}{m^5_q}\right).\label{25}\ee

Comparing  the expansion of $( m^2_q+\Delta m^2_q- D^2_\mu)^{-1}$
with (\ref{10}) one can define the QSE for $m_q\to 0$
\be
\Delta m^2_q=-\Lambda_0=-\frac{1}{\delta^2} (4D(0)+\frac32D_1(0))
.\label{26}\ee For purely exponential correlator $D(x)$ one can
connect $D(0)$ with string tension $\sigma$; in the Gaussian
approximation, which was checked recently using Casimir scaling
arguments to be accurate within few percents \cite{18,19}, one has
\be
\sigma=\frac12\int D(x) d^2 x=\frac{\pi
D(0)}{\delta^2}.\label{27}\ee As a result one has for the
self-energy if $m_q\to 0$
\be
\Delta
m^2_q=-\frac{4\sigma}{\pi}(1+\xi),~~\xi=\frac38\frac{D_1(0)}{D(0)}.\label{28}\ee
Lattice measurements \cite{16} give $\xi<\frac18$, and later we
shall omit this term. We note at this point that Eq.(\ref{28}) is
universal in the sense that it does not depend on the exponential
or  any other form of correlator in (\ref{15}), since the integral
(\ref{18}) for $m_q\to 0$ assumes the form of integral (\ref{27})
for string tension.

For nonzero values of $m_q$ one has instead of (\ref{28}) \be
\Delta m^2_q(m_q)=-\frac{4\sigma}{\pi}\eta,~~ {\rm with}~~
\eta=\frac{\varphi(m_q,\delta)}{\varphi(0,\delta)}.\label{29}\ee

For example, for the strange quark with  $m_s=0.175$ GeV one has
$\eta\cong$ 0.88, while for the $c$-quark, $m_c\cong1.7$ GeV,
$\eta\cong 0.234$ and finally for the $b$-quark, with $m_b\approx
5$ GeV one obtains $\eta\cong 0.052$.

\section{Self-energy corrections to Hamiltonian and hadron masses}

Using expression (\ref{5}) for the meson Green's function and
exploiting the einbein formalism \cite{9,10} one can construct the
relativistic Hamiltonian. We introduce the squared quark masses
with corrections due to QSE obtained in the previous section,
$m^2_q \to m^2_q+\Delta m^2_q$ and arrive at the expression $$
H=H_0+\Delta H_Q+\Delta H_s,$$ $$
H_0=\sum^2_{i=1}\left(\frac{m^2_q(i)}{2\mu_i}+
\frac{\mu_i}{2}\right)+\frac{p^2_r}{2\tilde \mu} + \frac{\hat
L^2/r^2}{2[\mu_1(1-\zeta)^2+\mu_2\zeta^2+\int^1_0 d\beta
(\beta-\zeta)^2\nu(\beta)]}$$
\be
+\frac{\sigma^2
r^2}{2}\int^1_0\frac{d\beta}{\nu(\beta)}+\int^1_0\frac{\nu(\beta)}{2}
d\beta\label{30}\ee  and \be \Delta H_q=\sum^2_{i=1}\frac{\Delta
m^2_q(i)}{2\mu_i}.\label{31} \ee

Here we have defined the reduced dynamical mass $\tilde \mu$ and
the parameter $\zeta$: \be\tilde
\mu=\frac{\mu_1\mu_2}{\mu_1+\mu_2},~~ \zeta
=\frac{\mu_1+\int^1_0\beta \nu (\beta)
d\beta}{\mu_1+\mu_2+\int^1_0\nu d\beta}\label{32}\ee and $m_q(1)$
and $m_q(2)$ are current quark and antiquark masses, renormalized
at the scale around 1 GeV, while $\mu_1, \mu_2$ and $\nu(\beta)$
are einbein functions which should be found from the minimum of
this Hamiltonian. For the discussion and practical use of this
Hamiltonian including spin-depending part $\Delta H_s$, beyond the
original papers \cite{10} see also lectures \cite{11} and recent
paper \cite{12}.

It is important that the dynamical masses $\mu_1,\mu_2$ are found
from the stationary point analysis of the Hamiltonian and these
values $\mu_1^{(0)}, \mu_2^{(0)}$ can be used to estimate self-
energy corrections to the Hamiltonian and its eigenvalues.

Hence the eigenvalues of $H$ will be shifted by an amount read off
from (\ref{31})
\be
\Delta H_q^{meson}=-\frac{2\sigma}{\pi}\left(
\frac{1}{\mu_1^{(0)}}+\frac{1}{\mu_2^{(0)}}\right),\label{33}\ee
i.e. from each quark the shift is equal approximately to
$-\frac{2\sigma}{\pi\mu_1^{(0)}} $. Taking into account
 that the lowest eigenvalue $M^{(0)}$ is equal (without
color  Coulomb interaction) $4\mu^{(0)}$ \cite{9,10,12}, one has
for equal masses $\mu_1^{(0)}=\mu_2^{(0)}=\mu^{(0)}$
\be
M^{(0)}(L=n_r=0){+\Delta
M}=4\mu^{(0)}-\frac{4\sigma}{\pi\mu^{(0)}}.\label{35a}\ee

In a similar way one can write the shift of the baryon mass. Using
\cite{20,21} one can write
\be
\Delta H_q(baryon)
=-\frac{2\sigma}{\pi}\sum^3_{\i=1}\frac{1}{\mu^{(0)}_i},\label{34}\ee
where $\mu_i^{(0)}$ are to be found again from the minimum of
baryon Hamiltonian \cite{20,21}. In the lowest approximation
(color Coulomb and spin interaction neglected) and for light
quarks one has $\mu^{(0)}_i=\mu^{(0)}$ and $M^{(0)} =6 \mu^{(0)}$,
so that the resulting corrected mass is \be M=M^{(0}+\Delta H_q=
6\mu^{(0)}-\frac{6\sigma}{\pi\mu^{(0)}}.\label{35}\ee

The same type of correction occurs for hybrids, since $\Delta H_q$
has the same form (\ref{33}) as for mesons, however the values of
$\mu^{(0)}_1$ and $\mu^{(0)}_2$ are  to be found for the total
hybrid Hamiltonian (see \cite{11}) containing in addition
dynamical mass $\mu_g$ of the valence gluon.

We now come to the one of the most important problems of hadron
spectrum, which  to the knowledge of the author had before no
solutions -- the problem of Regge intercepts. It was shown in
\cite{10,24,25} that the Hamiltonian $H_0$  ensures the string
Regge slope and the square of its mass eigenvalue is
$M^2(L)=2\pi\sigma L+C_0$, where  $C_0$ is positive and  too large
(for $m_q=0$, $C_0\approx 11\sigma =(1.4$ GeV$)^2$ when Coulomb
and hyperfine interaction is not taken into account), hence the
Regge intercept
\be
J(M=0)=L+1= 1-\frac{C_0}{2\pi\sigma}\label{40}\ee is negative
(-0.73), while for the experimental $\rho$  trajectory it is
positive and near +0.5. Therefore one  needs such mass corrections
which would shift Regge trajectories upwards in the Chew-Frautschi
plot  $(J(M^2)$ vs $M^2)$ not changing its slope.

As we shall see now, the correction $\Delta H_q $ (\ref{30}) has
exactly this property. To start we take the corrected mass
(\ref{35})  and consider its square
\be
M^2\equiv(M^{(0)}+\Delta M)^2=
(4\mu^{(0)}-\frac{4\sigma}{\pi\mu^{(0)}})^2=(M^{(0)})^2-\frac{32\sigma}{\pi}
+O(\Delta M)^2)\label{41}\ee

Thus to our accuracy (first order in $\Delta M)$ mass squared is
shifted by a constant not depending on quantum numbers,
$-\frac{32\sigma}{\pi}$, and if $(M^{(0)})^2$ corresponded to a
linear Regge trajectory, the same will be true for the shifted
trajectory with the same slope.

If one calculates $M^{(0)} =4\mu^{(0)}$ from the simplified
Hamiltonian, obtained from (\ref{31}) by putting $\hat L\equiv 0$
and $p^2_r\to \vep^2$(this is very close to the relativistic quark
model Hamiltonian $H_{RQM}\equiv  2\sqrt{\vep^2+m^2_q}+\sigma r$)
then the slope of $M^{(0)}$ would be $8\sigma$ instead of expected
string-like $2\pi\sigma$ \cite{9}, and this slope is kept intact
by the shift (\ref{41}). To resolve this inconsistency  one should
take into account the term with $\hat L^2$ in (\ref{31}) which
contains  the moment of intertia of the rotating string and
changes the slope to the correct value $2\pi\sigma$ (\ref{11}).

Exact calculations with the Hamiltonian (\ref{31}) require
numerical solutions yielding the linear Regge trajectories with
the slope very close to $2\pi \sigma$ (see \cite{24,25} for
results and discussions). Below we shall consider the effect of
string rotation as a correction $\Delta M^{(1)} $ \cite{10} and
write for the whole mass squared
\be
M^2\cong (4\mu^{(0)}+\frac{\Delta m^2_q}{\mu^{(0)}}+\Delta
M^{(1)})^2\label{42}\ee with $\Delta m^2_q=-\frac{4\sigma}{\pi}$
and $ \Delta M^{(1)}=-\frac{32}{3}
\frac{\sigma^2L(L+1)}{(M^{(0)})^3}$. Denoting
\be
(4\mu^{(0)}+\frac{\Delta m^2_q}{\mu^{(0)}})^2=8\sigma L +\bar C_0,
\label{43}\ee
 one obtains to the lowest order
\be
M^2\cong 8\sigma L+\bar C_0-\frac{32}{3}
\frac{\sigma^2L(L+1)}{8\sigma L+\bar C_0}.\label{44}\ee

The Regge slope $\frac{\partial M^2}{\partial L}$ can be computed
from (\ref{44}) both for $L=1$ and for $L\to\infty$, which yields
$\frac{\partial M^2}{\partial L}=\frac{20}{3}\sigma (L\to
\infty)$, $\frac{\partial M^2}{\partial L}\cong 6.5 \sigma (L=1)$.

One can see that both  slopes are rather close to the string one,
$2\pi\sigma$, while  the self-energy term $\Delta m^2_q$
contributes only to the constant $\bar C_0$, shifting it towards
the experimental value of $m^2_\rho$. Thus QSE indeed can solve
the problem of Regge intercepts to a reasonable accuracy. To get
an idea of the magnitude of $\Delta H_q$, consider the $\rho$ -
meson, corresponding to the solution of the Hamiltonian (\ref{30})
with $L=n_r=0$. Taking $\sigma=0.18$ GeV$^2$ one obtains
\cite{9,11} $\mu_1^{(0)}=\mu^{(0)}_2=\mu^{(0)}=0.352$ GeV and
$M^{(0)}= 4\mu^{(0)}=1.41$ Gev, while (\ref{33}) yields  $\Delta
H_q(meson) =-0.65 $ GeV, with the total mass $M=M^{(0)} +\Delta
H_q= 0.76$ GeV. This last figure is close to the experimental
$\rho$-meson mass, but our result should be corrected by hyperfine
spin term and Coulomb interaction which tend mass in different
directions with a net shift around or less than 0.1 GeV.

A similar situation occurs for baryons. Taking $\sigma =0.15$
GeV$^2$ (see \cite{22} and \cite{23} for the phenomenologically
motivated choice and its theoretical justification), one obtains
from  \cite{21}
 for light quarks $\mu^{(0)}= 0.957 \sqrt{\sigma}$, and $M^{(0)}_B=6\mu^{(0)}$.
As a result one obtains for $\Delta$ mass (without spin and
Coulomb correction) $$M_B^{(0)}+\Delta M= (2.22-0.77) {\rm GeV}
=1.44 {\rm GeV}.$$ For strange baryon, $\Omega$, one has instead
for $m_q(s)=0.175$ GeV the dynamical mass $ \mu_s=$0.415 GeV
\cite{21} and finally $$M^{(0)}(\Omega)+\Delta H_q=(2.46-0.607)
{\rm GeV}=1.85 {\rm GeV}.$$

These examples are only to illustrate the order of magnitude of
resulting corrections. We leave the problem of realistic meson and
baryon mass calculations with spin and Coulomb interaction
included to future publications.

\section{ Discussion and conclusions }

In most model calculations of hadron masses done  heretofore the
absolute values of masses have been  defined  up to a  constant,
varying from one family of hadrons to another. Also in our
previous calculations \cite{11,12} this constant was introduced
and associated with the quark self-energies, however the latter
were defined phenomenologically \cite{12} to be around -0.25 GeV
per light quark, but not derived from theory.

In the present paper we have looked into problem of quark
self-energies and absolute values of hadron masses, and calculated
both  in terms of the only scale of nonperturbative QCD - the
string tension $\sigma$.

As a result we have succeeded in several respects. Firstly, all
our theory is based on quark current masses, assumed to be
renormalized at the scale of around 1 GeV, where NP effects do not
yet practically  enter. These masses should be associated with the
pole masses, introduced in case of heavy quarks (see e.g.
\cite{13}).

Secondly, QSE corrections yield a numerically reasonable shift of
masses downwards, and  for light quark this shift is indeed around
-0.25 to -0.3 GeV, as was required by phenomenology \cite{12}.

 Thirdly, and this is most important  featurefrom theoretical point of
 view, this QSE shift  does not spoil the correct Regge slope and
 at the same time resolves the old problem of Regge intercept which was too
  low in all previous calculations
 (if negative constant is not introduced).

 Let us now discuss the nature of this correction and compare it
 to another possible approach. First of all, the negative sign of
 QSE is evidently connected to the paramagnetic mechanism, which is
 at work here.
 One should note that the quark charge interaction, present in
 (\ref{4}) in the form of $gA_\mu$, produces effect of another
 (positive) sign to the quark mass squared, and the net result is
 bounded from below, since in $(m^2-\hat D^2)^{-1}$ the operator
 $-\hat D^2 =-D^2_\mu - g(\sigma F)$ is nonnegatively defined, and
 vanishes only for quark zero modes.
 However  in the confining phase the charge part $gA_\mu$ produces
 for the white system of  ($q\bar q)$ the confining interaction
 $\sigma r$ and belongs to the whole worldsheet of the  string and
 cannot be associated with QSE. Instead the  part $g(\sigma F)$
 contains the local correction (in the limit of small $T_g$)
 $g^2\lan (\sigma F)(\sigma F)\ran,$
 which can be treated as QSE and was computed in the present
 paper. One important conclusion is that actually QSE should be
 considered finally as a correction to the total hadron mass, and
 not to the mass of the quark, since the latter has no definite
 meaning inside the hadron. Therefore one should not give much
 sense to the fact that for light quarks the total quark mass
 $m^2_q+\Delta m^2_q$ is negative --one should have in mind that
 this contribution is only a part of the Hamiltonian (\ref{30})
 and should be computed as a correction to its eigenvalues. (Note
 that one cannot include $\Delta H_q$ in $H_0$ to find
 $\mu_i^{(0)}$ from stationary point analysis, since this
 procedure would go beyond the lowest order in $\Delta H_q$, while
  $\Delta H_q$ was itself computed to the lowest order only).

  To get more understanding of negative QSE correction to the
  quadratic (in Dirac operator $\hat D$) Hamiltonian, one can
  consider masses of heavy-light mesons $M_{HL}=M_Q+\varepsilon$
  computed in two different approaches, namely to
  compare eigenvalues of this quadratic Hamiltonian  in the case when  one mass,
  $m_q^{(2)}\equiv M_Q$, is  very heavy, to the eigenvalues of the corresponding
  Dirac equation with linear confining interaction.  Taking the
  latter from \cite{26,27} one has for lowest Dirac eigenvalues,
  \be
  \varepsilon_0^{(D)}= 1.62 \sqrt{\sigma}
  \label{45}
  \ee
  while for the Hamiltonian $H_0(m^{(2)}_q\to \infty)$ one obtains
  \cite{12}
\be
  \varepsilon_0^{(H)}= 2.26 \sqrt{\sigma}.
  \label{46}
  \ee

One can see the difference of about 0.3 GeV, which is very close
to the  mass shift due to QSE. This is not surprising since in
$\varepsilon_0^{(D)}$  the corresponding QSE correction   is
absent, while for $\varepsilon_0^{(H)}$ this correction is present
and is given by  $\frac{\Delta m^2_q}{2\mu}$. Therefore working in
the linear Dirac formalism like that developed in \cite{28} one
should not include the aforementioned QSE correction and the
Regge-intercept problem does not appear.
  Finally it is interesting to note, that self-energy correction
  similar to the one considered in the present paper appears in
  the 1+1 QCD, where in 't Hooft equation \cite{32} each quark
  obtains a negative contribution to the mass squared equal to
  $-2\sigma/\pi$, i.e. one half of self-energy correction
  computed above in 3+1 QCD.

Summarizing, we have calculated the nonperturbative QSE correction
explicitly and found the shift in hadron masses and Regge
intercept which is strongly favored phenomenologically.

Results of more detailed analysis for both light and heavy hadrons
are to be published elsewhere.

The author is greatly indebted to A.M.Badalian and A.B.Kaidalov
for many stimulating discussions. This work was financially
supported by the RFFI grants 00-02-17836 and 0015-96786.

\end{document}